\documentstyle[11pt,newpasp,twoside,epsf]{article}
\begin{document}
\title{ATCA and CMB anisotropies}
\author{Ravi Subrahmanyan}
\affil{Australia Telescope National Facility, CSIRO,
	Locked bag 194, Narrabri, NSW 2390, Australia}

\begin{abstract}
Australia Telescope Compact Array (ATCA) observations
in sky regions selected to be low in foreground confusion
have been used to infer limits on arcmin-scale CMB anisotropy
in total intensity and polarization.  These deep searches have, hitherto,
been made using the East-West ATCA in an ultra-compact 
1-D array configuration
and in the highest available frequency band at 3-cm wavelength.
The ATCA is being upgraded for operation at mm wavelengths and 
in 2-D array configurations:
the enhanced capabilities present a new opportunity for pursuing
CMB anisotropy related observations with reduced foreground confusion
and at smaller angular scales.
\end{abstract}

\section{Introduction}
Observations of primary cosmic microwave background (CMB) 
anisotropies and estimates of the angular spectrum
of the anisotropies, when compared with predictions for these spectra
based on structure formation theory and the propagation of the CMB 
photons through growing matter perturbations in the recombination era
and at later times, have long been expected to 
provide us with useful estimates of the 
cosmological parameters and the primordial perturbation 
spectrum (see, for example, Bond, Efstathiou, \& Tegmark 1997).
In a spherical harmonic decomposition of the CMB anisotropy, the primary
anisotropy
spectrum is expected to be exponentially cut off at multipole orders $l$
exceeding about 2500 (Hu \& White 1997).  
At higher multipoles --- at arcmin and smaller
angular scales --- the CMB spectrum is expected to be dominated by
secondary effects: anisotropy owing
to the Sunyaev-Zeldovich effect (SZE; see, for example, 
Refregier et al. 2000 and references therein) in a cosmological distribution 
of hot gas in cluster potential wells
is predicted to be the most important contributor.  The CMB is also
expected to be fractionally polarized --- the polarization has 
to date not been detected --- and any measurements of the angular spectrum
in polarization modes or the correlation between polarization
anisotropy and that in total intensity would provide additional
constraints on cosmology and help resolve certain parameter 
degeneracies (Seljak \& Zaldarriaga 1997).

\section{Australia Telescope CMB results to date}

The Australia Telescope Compact Array (ATCA; see The Australia Telescope
1992) is located at Narrabri in Australia, has five 22-m diameter 
antennae movable along a 3-km East-West railtrack, a sixth 
antenna stationed 3~km from the end of the track,
and usually operates
as an interferometric Fourier Synthesis array.  The array was commissioned 
with 20, 13, 6 and 3-cm capability and has been
in operation for more than a decade.  

CMB anisotropy searches using the ATCA have, so far, been made 
at the highest available frequency band of
3-cm where foreground confusion is lowest.  To maximize the array
brightness sensitivity, the five antennas were close-packed into
a 1-D East-West 122-m array in which the antennas were 
equi-spaced 30.6~m apart.  Shadowing results in spurious responses 
at the ATCA (Subrahmanyan 2002) and, therefore, fields were selected 
at declinations close to $-50^{\circ}$ to avoid shadowing.
At this declination, the projected antenna spacing is close to the
antenna diameter for a large hour angle range and observations
at this declination have the best brightness sensitivity (without shadowing).

Fields were observed in full Earth-rotation
Fourier synthesis mode and the strategy has been to examine images 
of the fields synthesized from the visibility data for any CMB anisotropy.
The five-antenna 122-m array gives 30.6-m
spacing visibilities on four baselines; the remaining six baselines
have spacings from 61 to 122~m.  Images synthesized using exclusively 
the four 30.6-m baselines have the highest brightness sensitivity:
these images are the most sensitive to flat-band CMB anisotropy.
The longer baselines are relatively insensitive to flat-band
anisotropies; however, they respond to foreground discrete
sources which are unresolved on those baselines.
Therefore, the strategy has been to use images made using 
the six longer baseline data
to estimate the foreground discrete source confusion in the fields
and use images made using the four 30.6-m baselines for CMB anisotropy
searches.  Stokes I, Q \& U images were examined for 
any CMB anisotropy.  The Stokes V images served as an estimator of the
instrument thermal noise.

In Subrahmanyan et al. (1993) were reported results based on observations
of a single field at 8.7~GHz.  This was followed by a
deeper integration on the single field, 
along with observations of the foreground confusion at a lower
frequency of 4.8~GHz in a roughly scaled array in which the five antennas
were equi-spaced 61-m apart to form a 244-m array (Subrahmanyan et al. 1998).

Results from deep observations of six separate fields at 8.7~GHz 
are in Subrahmanyan et al. (2000).  These six fields had been selected to
be low in foreground discrete sources based on lower frequency ATCA
observations.  The fields were observed for about 
50~hr each and the 2-arcmin resolution images, which were made using
exclusively the 30.6-m baselines, had rms thermal noise 
of 24~$\mu$Jy~beam$^{-1}$: this was as expected given the observing
parameters and telescope system temperatures.  
The telescope filter function, corresponding to the CMB sky variance as
estimated from primary-beam weighted image pixels, peaks at multipole
$l=4900$ and has half-maximum values at $l=3520$ and 6200.  
Each of the six 2-arcmin resolution images 
has about $2 \pi$ independent pixels and the six
images together represent $12 \pi$ independent estimates of the CMB sky
as viewed through this telescope filter function.  A flat-band
CMB anisotropy spectrum with normalization $Q_{flat}$~$\mu$K is expected to
result in an image rms $0.844Q_{flat}$~$\mu$Jy~beam$^{-1}$; consequently,
the image thermal noise corresponds to 28~$\mu$K.

\subsection{Results on polarization anisotropy}

The Stokes Q and U images of all six fields were consistent with
the expectations from thermal noise. The image rms, estimated using 
pixels in all the six images with appropriate weights, was 22.2
and 21.8~$\mu$Jy~beam$^{-1}$ in Stokes Q and U respectively.  
No polarization anisotropy is detected and, using the likelihood
ratio test, upper limits of $Q_{flat} < 11$ and 10~$\mu$K are derived, 
respectively, for Stokes Q and U CMB anisotropy in the ATCA filter function.

\subsection{Results for total intensity anisotropy}

The rms of the sky intensity fluctuations in Stokes~I, 
estimated using all the six images, was 
117~$\mu$Jy~beam$^{-1}$; there was significant `excess' image variance
above that expected from the telescope thermal noise.  The higher
resolution images made using visibilities with baselines in the 61-122~m
range showed obvious foreground discrete sources in the fields.
The foreground confusion was modelled using the long baseline data ---
the model derived consisted of 0-4 `point' sources in each field
with flux density up to 455~$\mu$Jy~beam$^{-1}$ --- and the model
was subtracted.  The image rms derived from the residual images was
52~$\mu$Jy~beam$^{-1}$ and this was also significantly 
in excess of the thermal noise.

The residual confusion in the residual images was estimated via
simulations. The differential source counts at 8.7~GHz, derived
from deep VLA observations by Windhorst et al. (1993), was used
to generate simulated sky images containing the expected thermal
noise and discrete sources. The images were processed in a pipeline
which simulated the observing strategy and data processing including
the identification of confusion from the long-baseline data.
The simulated residual images were used to derive the distribution
for the estimate of the residual image rms. The simulations suggested
that the residual weak-source confusion may cause 
the residual images to have rms 51-66~$\mu$Jy~beam$^{-1}$ (1-$\sigma$ spread). 
This estimate is consistent with the observational results and, therefore,
we have been led to believe that the `excess' variance in the residual
images, following the subtraction of the confusion model, is owing to
weak unsubtracted confusion and not CMB anisotropy.  

Assuming that thermal noise and discrete source confusion (including
weak confusion from sources not identified in the long-baseline 
data) are the only contributors to the image apart from any flat band
CMB anisotropy, a likelihood ratio test gave upper limits 
of $Q_{flat} < 25$~$\mu$Jy~beam$^{-1}$ with 95 per cent confidence.

\section{ATCA mm upgrade} 

The ATCA is being upgraded for mm operations.  All five 
movable antennae are
being fitted with 12 and 3~mm receivers covering
the bands 16-25 and 85-115~GHz.  The receivers are being built around
Indium Phosphide MMIC low-noise amplifiers which are designed in-house
at the Australia Telescope National Facility (ATNF) laboratories
in Sydney.  Prototype receivers installed on
two antennae have receiver temperatures (measured at the face of the feed
horns) of about 25 and 100~K respectively at 21 and 95~GHz; 
the system temperatures
are 45 and 175~K at these frequencies.  Temperatures in the 3~mm band are 
expected to improve in later versions.

The antenna surfaces have been upgraded for mm operations: panels with
perforations have been replaced with solid panels which have $< 100$~$\mu$m
rms errors and holography has been used to reset the panels.
The receiver feed horns illuminate
the entire 22-m aperture of the antennae even at 3~mm: the ATCA 
antennas have shaped reflectors for the Cassegrain optics and the aperture
illumination is fairly uniform out to the edges and this gives the
short-baseline interferometers good brightness sensitivity.  Because the
peripheral regions of the
antenna surfaces are important for brightness sensitivity, the 
holographic panel setting used a feed horn which over illuminated the
edges and improved sensitivity for locating the outer panels.
All these imply that the ATCA has good brightness sensitivity and this is
important for CMB related work.

The atmosphere is
a significant degrader of system temperature particularly at mm
wavelengths; therefore, observations
are better done at high elevations.  In order to have good
visibility-plane coverage from observations at exclusively
high elevations and short durations, the upgrade includes the
addition of a North-spur to the current East-West railtrack.
CMB observations with the ATCA have, in the past, been made with 
1D close-packed arrays.
The brightness sensitivity of an interferometric array is better
if the array configuration is close-packed in 2D.  The upgrade
makes it possible for CMB observations with the ATCA, in future, 
to be made in 2D configurations.

A separate project, which is also part of the upgrade, is 
a redesign of the
phase transfer system to provide phase-stable reference local oscillators
at each antenna of the interferometer array
for down-conversion of the high frequencies without loss of coherence.

A related development, which is expected to come off in the next few years,
is an upgrade of the ATCA correlator to wider bandwidths. The increase
from the current $2 \times 128$~MHz to several GHz will vastly enhance
the continuum sensitivity of the interferometer: this is vital for
CMB anisotropy related observations where sky signals are extremely weak.

\section{Prospects for ATCA CMB observations in the future}

Many CMB anisotropy related observations are now being made at
mm wavelengths, about 100~GHz, where confusion from synchrotron
and dust foregrounds are believed to be lowest.  The focus is also
shifting to smaller angular scales and here the errors in the estimates
of the CMB power spectrum may be dominated by discrete source
confusion.  However, all estimates of source counts at mm wavelengths
are based on mm observations of sources detected in cm wavelength
surveys: there are no surveys which cover 
significant sky areas at mm wavelengths.
The 3-mm capability in the upgraded ATCA may make a useful contribution
in de-confusing CMB surveys.  It may be noted here 
that the region of the sky which
is lowest in dust emission is in the vicinity of the south Galactic pole,
in the southern sky, and this is where many current CMB anisotropy
searches are being directed.  The ATCA is the only mm interferometer
capable of observing this region today and the upgraded telescope
may play a role in detecting and monitoring the stronger discrete
sources in the CMB fields for de-confusing total intensity or
polarization anisotropy observations.

\begin{figure}
\plotfiddle{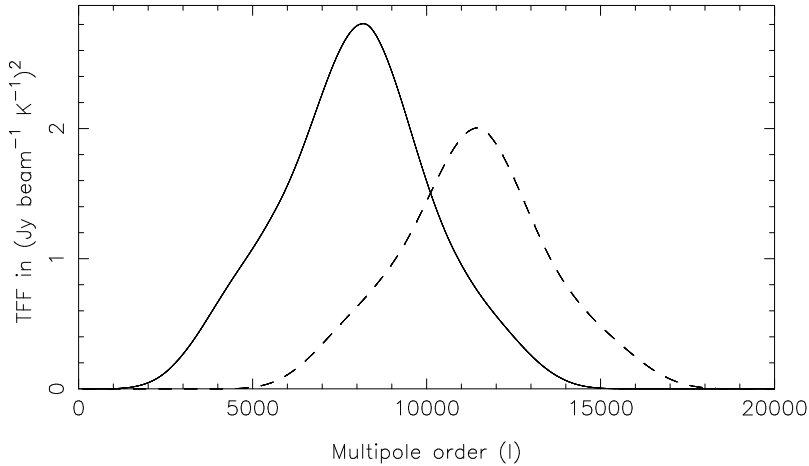}{6 truecm}{0.0}{60}{60}{-160}{-160}
\caption{Telescope filter functions for a single ATCA baseline
	at 18~GHz.  The continuous line is for a projected baseline of
	22~m; the dashed line is for 30.6~m.}
\end{figure}

In any ultra-compact array configuration, assuming that the system
temperature and other observing parameters remain the same, 
and for frequencies in the Rayleigh-Jeans regime 
($h \nu \ll k T_{\circ}$; where $T_{\circ} = 2.726$~K), the ATCA
sensitivity to flat-band CMB anisotropy does not change with the
observing frequency.  The interferometer response, in Jy~beam$^{-1}$,
is the same for a fixed $Q_{flat}$ or for a flat spectrum with
$l(l+1)C_{l}/(2 \pi)$ a constant.

However, primary anisotropies are believed to cut off beyond about
$l \sim 2500$ and SZE anisotropy searches with angular resolutions 
much better than an arcmin
are not currently considered useful; therefore, anisotropy searches
with the upgraded ATCA may be made at the lower end of the 12~mm band,
around 16-20~GHz.  The telescope filter function, corresponding to
an interferometer consisting of two 22-m ATCA antennas spaced 30.6~m apart,
is given in Figure~1.  The continuous curve gives the filter corresponding to
a projected baseline of 22~m, the dashed line gives the filter corresponding
to a projected baseline of 30.6~m; both curves have been computed assuming 
operation at 18~GHz.  The telescope response peaks at $l$ in the range
8000-12000 depending on the fore-shortening of the baseline in projection.
Assuming that at arcmin scales, the secondary
CMB anisotropy may be described by a flat band CMB anisotropy spectrum,
I find that for $\sqrt{l(l+1)C_{l}/(2 \pi)} 
\times 2.726$~K = 10~$\mu$K,
the expected signal is 10~$\mu$Jy~beam$^{-1}$
when the ATCA baseline is 30.6~m and 14~$\mu$Jy~beam$^{-1}$
when the projected baseline is 22~m.  

The extragalactic discrete source confusion would be very much
reduced at the higher frequencies available following the upgrade.
If we adopt a differential source
count which scales with the flux density $S$ and frequency $\nu$ as

$$N(S) \sim S^{-2.18} \nu^{-0.8 \alpha}, \eqno(1)$$

\noindent where $\alpha = 0.7$ is the spectral index (defined by the relation 
$S_{\nu} \sim \nu^{-\alpha}$), the integral source count obeys the
scaling relation:

$$N(>S) \sim S^{-1.18} \nu^{-0.56} \Omega, \eqno(2)$$

\noindent where $\Omega$ is the beam solid angle (which scales with
frequency as $\nu^{-2}$).  The confusion limit may be taken to be the
flux density $S_{c}$ at which $N(>S_{c}) \approx 1$; 
it follows that the confusion
limits scales with frequency as $S_{c} \sim \nu^{-2.17}$.  As compared
to previous ATCA CMB observations at 8.7~GHz, any future observations
at 18~GHz, with identical observing strategy, would have a confusion
that is lower by factor 5.

\acknowledgments
The Australia Telescope is funded by the Commonwealth of Australia for
operation as a National facility managed by CSIRO.

\end{document}